\def\be{\begin{equation}}
\def\ee{\end{equation}}
\def\beqn{\begin{eqnarray}}
\def\eeqn{\end{eqnarray}}

\def\ba{\begin{array}{c}}
\def\bat{\begin{array}{cc}}
\def\ea{\end{array}}
\def\bi{\begin{itemize}}
\def\ei{\end{itemize}}

\def\cR{{\cal R}}

\newcommand{\bel}[1]{\be\label{#1}}
\newcommand{\e}{\mbox{\rm e}}
\newcommand{\ta}[0]{\tilde \alpha}
\documentclass{PoS}

\usepackage{mathtools}
\usepackage{amsmath}
\usepackage{epstopdf}

\title{Constraining the two-Higgs doublet models with the LHC data}

\ShortTitle{Constraining the THDMs with the LHC data}

\author{\speaker{Victor Ilisie}\\
        Departament de F\'{\i}sica Te\`orica, IFIC,
        Universitat de Val\`encia -- CSIC,
        Apt. Correus 22085, E-46071 Val\`encia, Spain\\
        E-mail: \email{Victor.Ilisie@ific.uv.es}}

\abstract{The recent discovery of a Standard Model-like boson with  mass of about 126 GeV seems to be the first direct information on the electroweak symmetry breaking mechanism. Using the available experimental data from the LHC and Tevatron we study the implications on the parameter space of the Two-Higgs Doublet Model extension of the Standard Model. The generic structure of the Aligned Two-Higgs Doublet Model (ATHDM) is imposed in the Yukawa sector; also the models with discrete $\mathcal{Z}_2$ symmetries are analyzed.}

\FullConference{The European Physical Society Conference on High Energy Physics -EPS-HEP2013\\
		18-24 July 2013\\
		Stockholm, Sweden}

\begin{document}

\section{Introduction}

The recent discovery at the LHC of a boson with Standard Model (SM) Higgs-like properties brings us to a whole new horizon of possibilities regarding its nature and origin. Many theoretical models are able to reproduce the properties of this particle with mass of about 125 GeV and also generate interesting new physics at the TeV scale. The simplest extension of the SM, which has a richer scalar sector and that could give rise to new interesting phenomenology also in the flavour physics sector is the Two-Higgs-Doublet Model (THDM) \cite{Gunion:1989we,Pich:2009sp,Paper:2013rcs}. Next we shall analyze the phenomenology of the scalar sector of this model and see how the new LHC data constrains its parameter space.

\section{The Two-Higgs-Doublet Model}

The THDM extends the SM with a second scalar doublet of hypercharge $Y=\frac{1}{2}$.
The neutral components of the scalar doublets $\phi_a(x)$ ($a=1,2$) acquire vacuum expectation values that are, in general, complex: $\langle 0|\phi_a^T(x)|0\rangle =\frac{1}{\sqrt{2}}\, (0,v_a\, \e^{i\theta_a})$. Through an appropriate $U(1)_Y$ transformation we can enforce $\theta_1=0$, since only the relative phase $\theta \equiv \theta_2 - \theta_1$ is observable.
It is convenient to perform a global SU(2) transformation in the scalar space $(\phi_1,\phi_2)$ and work in the so-called Higgs basis
$(\Phi_1,\Phi_2)$, where only one doublet acquires a vacuum expectation value. In this basis, the two doublets are parametrized as
\begin{align}  \label{Higgsbasis}
\Phi_1 =\left[ \begin{array}{c} G^+ \\ \frac{1}{\sqrt{2}}\, (v+S_1+iG^0) \end{array} \right]  \!\  ,   & &
\Phi_2 = \left[ \begin{array}{c} H^+ \\ \frac{1}{\sqrt{2}}\, (S_2+iS_3)   \end{array}\right]  \!\ .
\end{align}
Thus, $\Phi_1$ plays the role of the SM scalar doublet with
$v\equiv \sqrt{v_1^2+v_2^2}\simeq (\sqrt{2}\, G_F)^{-1/2} = 246~\mathrm{GeV}$. The physical scalar spectrum contains five degrees of freedom: the two charged fields $H^\pm(x)$
and three neutral scalars $\varphi_i^0(x)=\{h(x),H(x),A(x)\}$, which are related with the $S_i$ fields
through an orthogonal transformation $\varphi^0_i(x)=\mathcal{R}_{ij} S_j(x)$.
The form of the $\mathcal{R}$ matrix is fixed by the scalar potential, which determines the neutral scalar mass matrix
and the corresponding mass eigenstates. A detailed discussion is given in \cite{Paper:2013rcs}. In general, the CP-odd component $S_3$ mixes with the CP-even fields
$S_{1,2}$ and the resulting mass eigenstates do not have a definite CP quantum number.
If the scalar potential is CP symmetric this admixture disappears; in this particular case, $A(x) = S_3(x)$
and
\bel{eq:CPC_mixing}
\left(\ba h\\ H\ea\right)\; = \;
%\left[\bat \cos{(\alpha - \beta)} & \sin{(\alpha - \beta)} \\ -\sin{(\alpha - \beta)} & \cos{(\alpha - \beta)}\ea\right]\;
\left[\bat \cos{\tilde\alpha} & \sin{\tilde\alpha} \\ -\sin{\tilde\alpha} & \cos{\tilde\alpha}\ea\right]\;
\left(\ba S_1\\ S_2\ea\right) \, .
\ee
Performing a phase redefinition of the neutral CP-even fields, we can fix the sign of $\sin{\ta}$.  In this work we adopt the conventions\ $M_h \le M_H$\ and\
$ 0 \leq \ta \leq \pi$ therefore, $\sin{\ta}$ is positive.

\section{Yukawa Alignment}

The most generic Yukawa Lagrangian with the SM fermionic content in the Higgs basis can be written as:
\begin{align}
\mathcal{L}_Y = -\frac{\sqrt{2}}{v}  \!\   \Big\{   \!\ 
\bar{Q}_L'  \!\ (M_d'\Phi_1 \!\ + \!\  Y_d'\Phi_2) \!\ d_R' \!\ + \!\ 
\bar{Q}_L' \!\  (M_u'\Phi_1 \!\ + \!\  Y_u'\Phi_2)u_R' \!\   
   +  \!\ \bar{L}_L' \!\  (M_l'\Phi_1 \!\ + \!\  Y_l'\Phi_2) \!\ l_R'  \!\ 
   \Big\} 
\end{align}
where ${Q}_L',  \!\ L_L \!\ (d_R', \!\  l_R')$ are the left-handed up type (right-handed down type) quark and lepton fields. $M_f'$ and $Y_f'$ ($f=u,d,l$) are the non diagonal mass and Yukawa matrices which are in general complex and independent; therefore these two matrices are not simultaneously diagonalizable. This gives rise to dangerous tree level flavour changing neutral currents (FCNCs) which are phenomenologically highly suppressed. In order to get rid of them one usually imposes a discrete $\mathcal{Z}_2$ symmetry on the Higgs doublets i.e., $\phi_1 \to \phi_1  \!\ , \!\  \!\ \phi_2 \to -\phi_2$ (in a generic basis), etc. However, a more general approach is to impose alignment in the flavour space $ Y_f' \sim  M_f'$ \cite{Pich:2009sp}. In terms of the the mass-eigenstate mass matrix we obtain
\begin{align}
&& Y_{d,l}=\varsigma_{d,l}  \!\ M_{d,l} \!\ , && Y_u=\varsigma_u^*  \!\ M_u \!\ ,&&
\end{align}
where $\varsigma_f$ ($f=u,d,l$) are called the alignment parameters. These three parameters are independent, flavour universal, scalar basis independent and in general complex. Their phases introduce new sources of CP-violation. The usual models based on $\mathcal{Z}_2$ symmetries are recovered taking the appropriate limits \cite{Pich:2009sp}. 
We can now write our Yukawa Lagrangian in terms of the mass-eigenstate fields:
\begin{align}
\mathcal L_Y  =  & - \frac{\sqrt{2}}{v}\; H^+  \Big\{  \bar{u} \Big[ \varsigma_d\, V M_d \mathcal P_R - \varsigma_u\, M_u^\dagger V \mathcal P_L \Big]  d\,  + \, \varsigma_l\, \bar{\nu} M_l \mathcal P_R l \Big\}
\nonumber \\
& -\,\frac{1}{v}\; \sum_{\varphi^0_i, f}\, y^{\varphi^0_i}_f\, \varphi^0_i  \; \left[\bar{f}\,  M_f \mathcal P_R  f\right]
\;  + \;\mathrm{h.c.} \notag
\end{align}
here $\mathcal P_{R,L}\equiv \frac{1\pm \gamma_5}{2}$ are the right-handed and left-handed chirality projectors and the  couplings of the neutral scalar fields are given by:
\begin{align}
\label{yukascal}
 y_{d,l}^{\varphi^0_i} = \cR_{i1} + (\cR_{i2} + i\,\cR_{i3})\,\varsigma_{d,l}   ,
&&
 y_u^{\varphi^0_i} = \cR_{i1} + (\cR_{i2} -i\,\cR_{i3}) \,\varsigma_{u}^* \, . 
\end{align}

\section{Phenomenology}

\begin{figure}[!htb]
\centering
\includegraphics[scale=0.57]{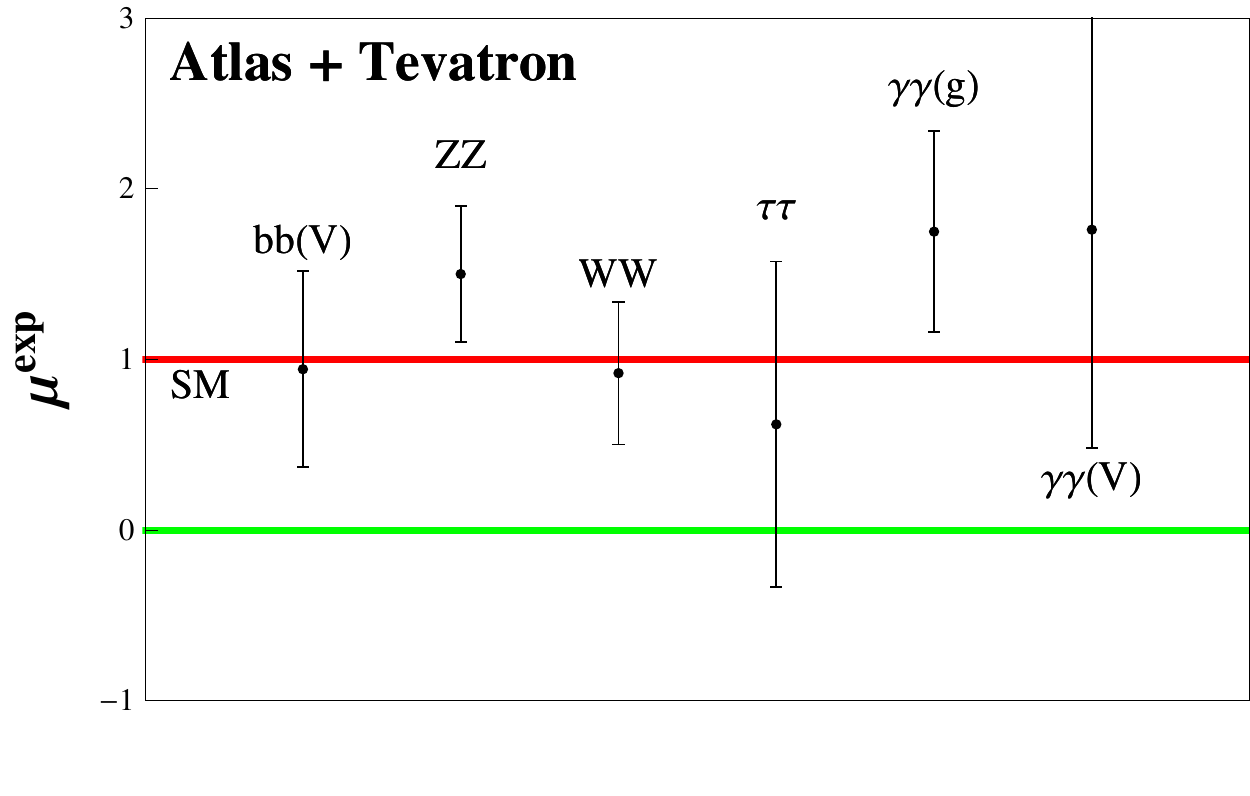}  \!\  \!\  \!\  \!\  \!\  \!\  \!\  \!\  \!\  \!\  \!\  \!\  \includegraphics[scale=0.57]{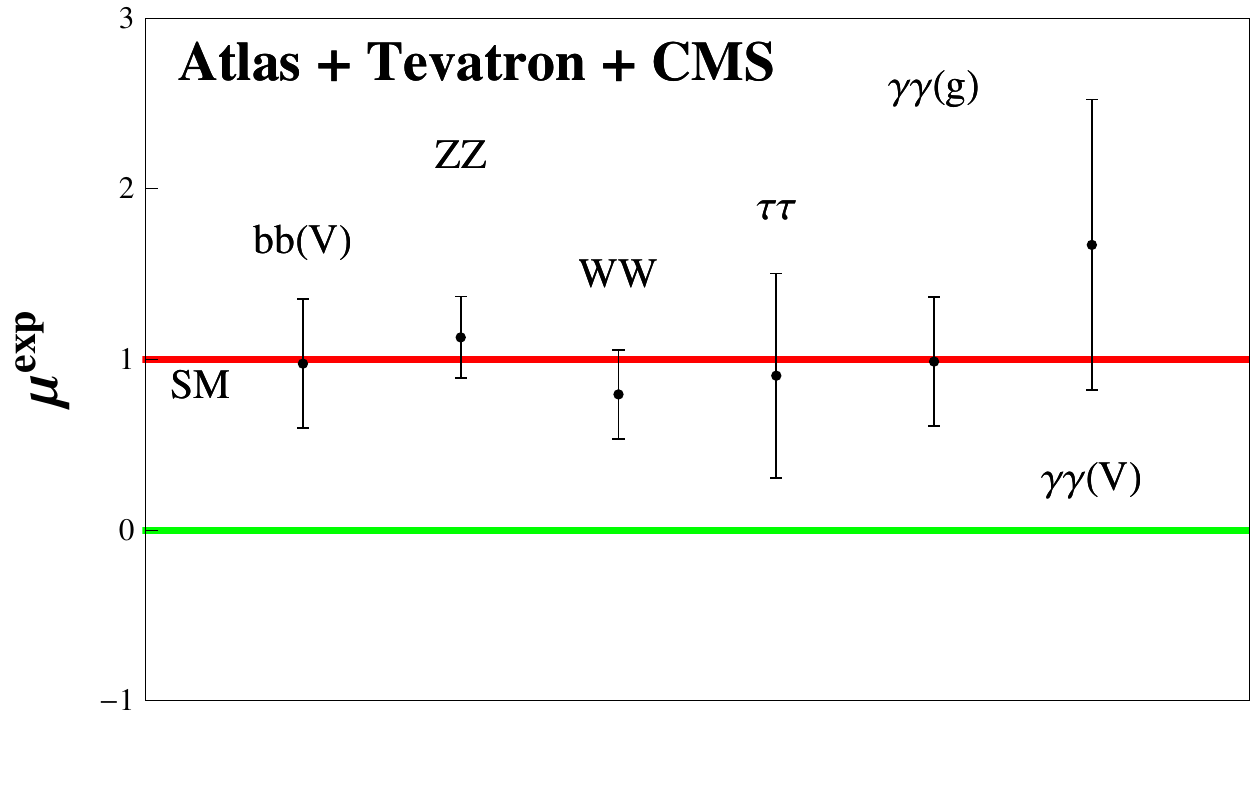}
\caption{Combined experimental data from Atlas and Tevatron (left) and including the CMS data (right)}
\label{expdata1}
\end{figure}

The latest experimental data provided by the ATLAS \cite{ATLAScombination} and CMS \cite{CMScombination} collaborations from the LHC together with the latest combined results from Tevatron \cite{Aaltonen:2013kxa} are in good agreement with the SM hypothesis, but the experimental errors are still large. However, one can use the present Higgs data to further constrain the correspondent parameter space of the theory. If one only combines the data given by ATLAS and Tevatron excluding the CMS contribution, one finds a slight excess for the center value of the two-photon decay channel both in gluon-fusion (ggF) and vector boson fusion (VBF) production, Fig.\ref{expdata1} (left). This excess is highly interesting because it might originate in new interesting physics. It could signal the presence of a charged Higgs (by adding an extra loop of a charged Higss to the $h\to \gamma \gamma$ decay) or it might originate in a different (than the SM), perhaps complex, Yukawa structure. If one includes the CMS data the excess is gone and all signal strengths are pushed closer to the SM, Fig.\ref{expdata1} (right). In the following we shall fit the parameter space to the experimental data with and without including the CMS results and compare the conclusions.

\subsection{Charged Higgs Contribution}

\begin{figure}[!htb]
\centering
\includegraphics[scale=0.46]{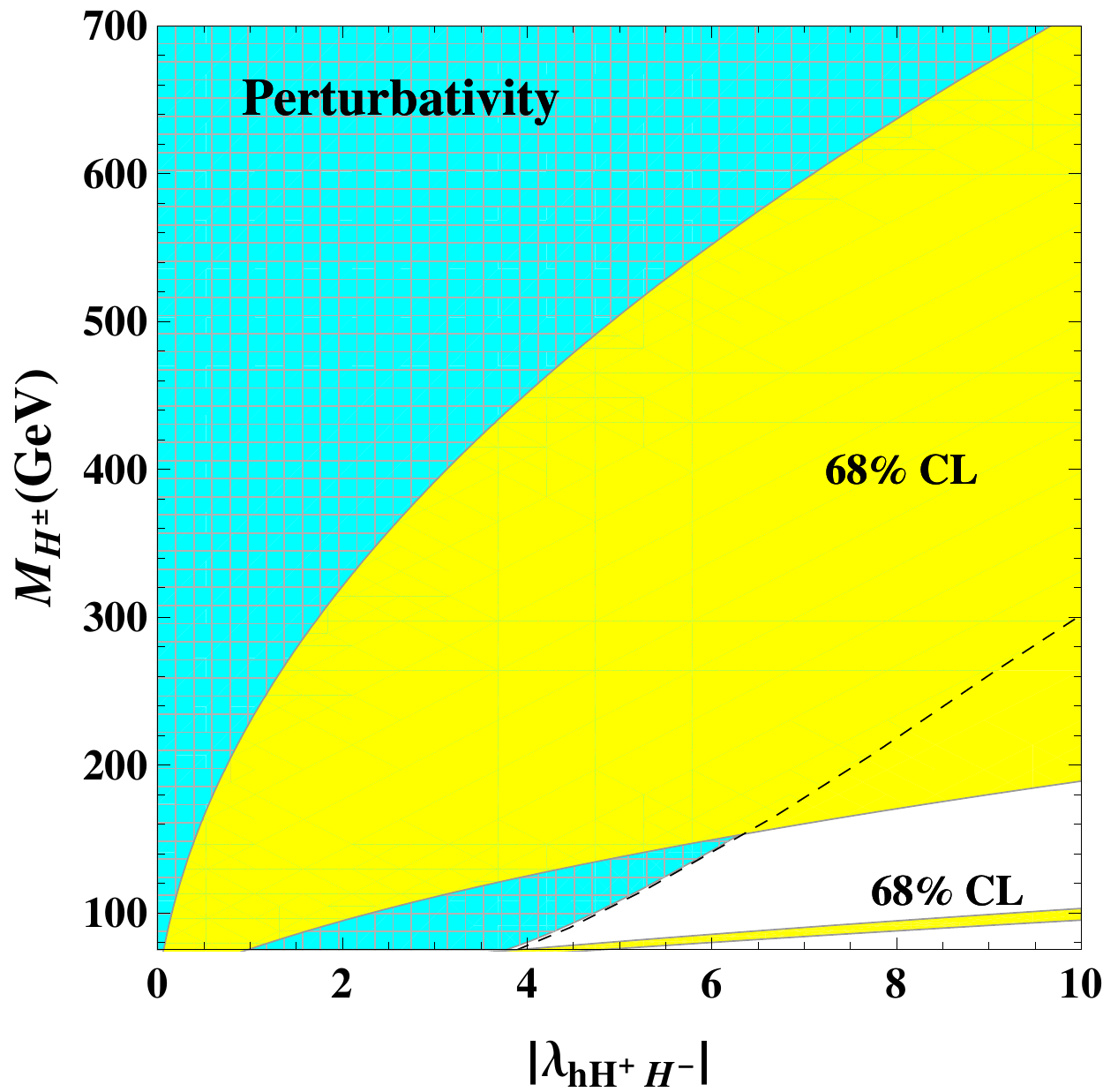}
\caption{One sigma allowed region for the ($|\lambda_{hH^+H^-}|,M_{H^\pm}$) parameter space (yellow, light). The perturbativity bounds are given by the blue (dark) area.}
\label{Chpm}
\end{figure}

In this section we assume that the Higgs potential and also the Yukawa sector are both CP-conserving; we shall also assume that the observed boson is the lightest CP-even scalar with $M_h=125$ GeV and that no other decays than the SM ones occur. In order to compute the $h\to \gamma\gamma$ decay width one needs the coupling of the neutral scalars to a pair of charged Higgs. Since these couplings depend on still
 unknown parameters we shall parametrize it as
\begin{align}
 \mathcal{L}_{hH^+H^-} = -  v \!\ \lambda_{hH^+H^-} \!\ h   H^+  H^-  \!\ .
\end{align}
For the first fit we only include the ATLAS and Tevatron data. The allowed one sigma region for the charged Higgs mass as a function of the coupling $\lambda_{hH^+H^-}$ is given in Fig.\ref{Chpm}. Perturbativity bounds \cite{Paper:2013rcs} have also been imposed. The two solutions correspond to either a constructive contribution of the $H^\pm$ and $W^\pm$ amplitudes (wide upper region) or to a destructive one but with a $H^\pm$ contribution so large that it reverses the sign of the total $h\to \gamma\gamma$ amplitude (narrow lower region). This last solution is excluded by the perturbativity bounds. The correspondent fitted Yukawas are given by
\begin{align}
\cos\ta = 0.98^{+0.2}_{-0.6} \; ,  &&  y^h_u = 1.0^{+0.4}_{-0.3} \; ,  &&  |y^h_d|=0.9 \pm 0.4 \; ,   &&    |y^h_l| = 0.7 \pm 0.6 \; , 
\end{align}

\noindent where $\cos\ta$ is the reduced coupling of $h$ to two massive gauge bosons. Changing simultaneously the signs of $\cos \tilde \alpha$ and $y_f^h$ leads obviously to identical Higgs signal strengths and, therefore, to equivalent solutions. There is a also a sign degeneracy in the bottom and tau Yukawa couplings. The only place where we are sensitive to the relative Yukawa signs is in the loop-induced processes, therefore the sign degeneracy is due to the fact that the dominating fermionic loop contribution in the $gg\to h$ and $h\to\gamma\gamma$ processes is the top quark contribution; all the other fermionic contributions are much smaller. 

When including the CMS data, the fitted Yukawas are simply given by
\begin{align}
\cos\ta = 0.98^{+0.2}_{-0.6} \!\ , & & y^h_u = 0.95\pm 0.25 \!\ , & & |y^h_d|=0.95 \pm 0.3 \!\ , & & |y^h_l| = 0.95 \pm 0.3  \!\ .
\end{align}
Since the contribution of the charged Higgs in this case is compatible with zero, we have not included it in the fit. As we can see all the center values are very close to the SM.

\subsection{ATHDM versus $\mathcal{Z}_2$ models}

\begin{figure}[!htb]
\centering
\includegraphics[scale=0.42]{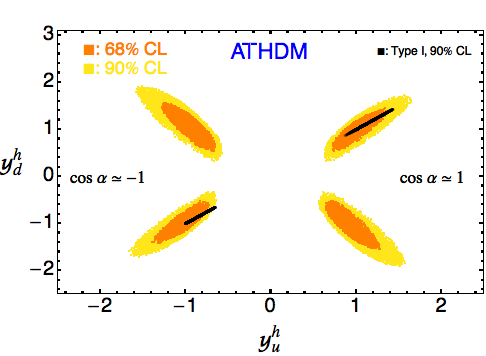}  \!\  \!\  \!\  \!\  \!\  \!\  \!\  \!\  \!\  \!\  \!\  \!\  \!\  \!\  \!\  \!\  \!\  \!\ 
\includegraphics[scale=0.42]{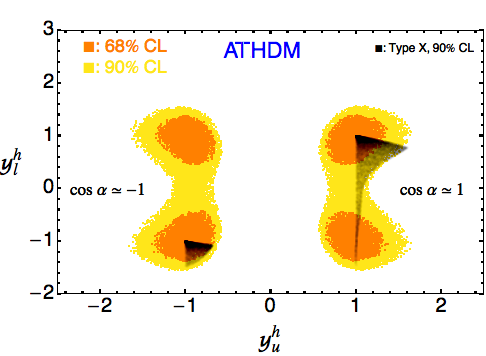}
\caption{One (orange, dark grey) and two (yellow, light grey) sigma allowed region for the parameter space of the ATHDM. The two sigma region for $\mathcal{Z}_2$ models are shown in black.}
\label{z2models}
\end{figure}

This section is dedicated to the comparison of the ATHDM, with CP-conseving Yukawas and potential, with the usual THDMs with $\mathcal{Z}_2$ symmetries \cite{Gunion:1989we,Nierste,Grinstein,Krawczyk,Cheng,Gunion,Val}. In this case we directly perform the fit including all data, thus the contribution from the charged Higgs is neglected for this analysis. In Fig.\ref{z2models} we can see the allowed region for the $(y_h^d,y_h^u)$ and $(y_h^l,y_h^u)$ parameter space. The allowed region for the ATHDM is still very wide and the $\mathcal{Z}_2$ type models are already very much constrained by the current experimental data. Again, in the two plots we can observe the sign degeneracy for the $y_l^h$ and $y_d^h$ couplings that we have encountered and explained in the previous section.

\subsection{Complex Yukawas}

\begin{figure}[!htb]
\centering
\includegraphics[scale=0.49]{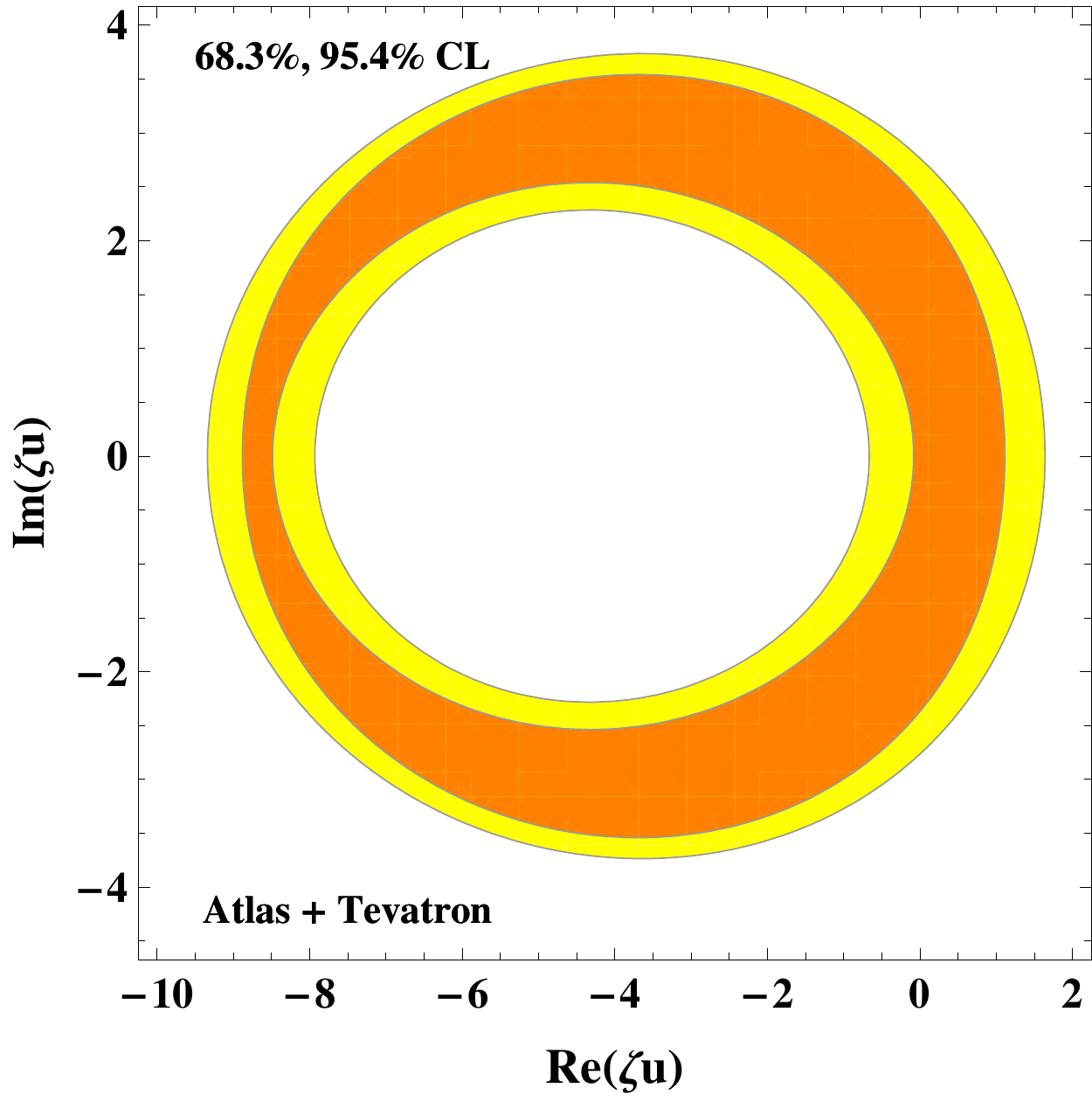}  \!\  \!\  \!\  \!\   \!\  \!\  \!\  \!\  \!\  \!\  \!\  \!\  \!\  \!\  \!\  \!\  \!\  \!\  \!\   \!\  \!\  \!\  \!\ 
\includegraphics[scale=0.49]{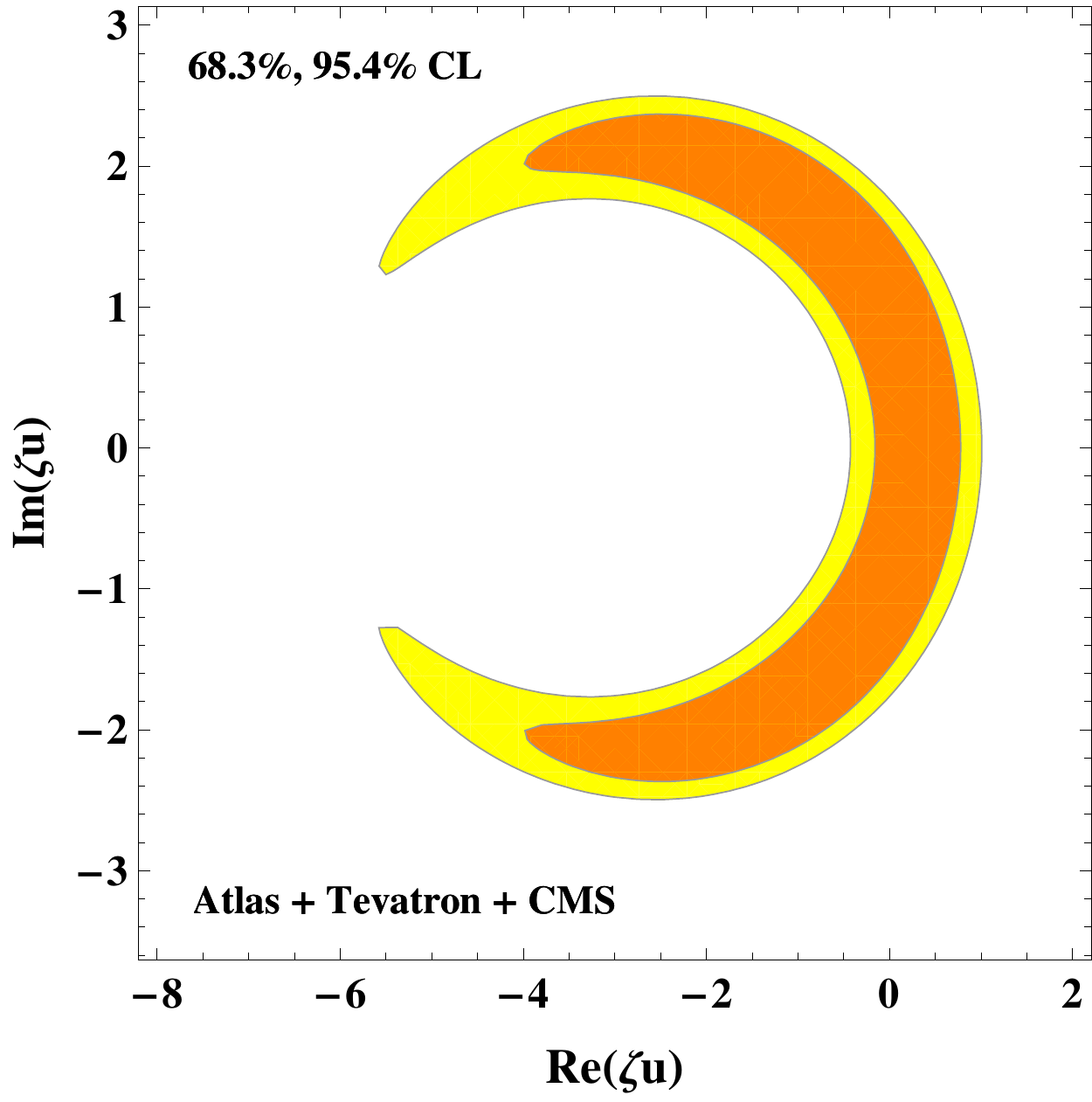}
\caption{One and two-sigma allowed region for the ($\text{Re}(\varsigma_u),\text{Im}(\varsigma_u)$) parameter space of the ATHDM.}
\label{complex1}
\end{figure}

For this last fit a CP-conserving potential is assumed but we allow a complex up-type Yukawa $y_u^h$; the other couplings $y_{d,l}^h$ are considered CP-conserving and their values are set to the best-fit point. When only ATLAS and Tevatron data are taken into account, the allowed one and two sigma regions for the real and imaginary part of $\varsigma_u$ are shown in Fig.\ref{complex1} (left). When the CMS data is also included, part of the allowed parameter space disappears, Fig.\ref{complex1} (right). The region that disappears contains the interval where the real part of $\varsigma_u$ is negative and high enough to flip the sign $y_u^h$ creating a positive interference between the $W$ boson and top quark loop. This way an excess in the $h\to \gamma\gamma$ is created. When including the CMS data, this region is forbidden because the excess in the signal strength of this channel vanishes.

\section{Conclusions}

\begin{figure}[!htb]
\centering
\includegraphics[scale=0.26]{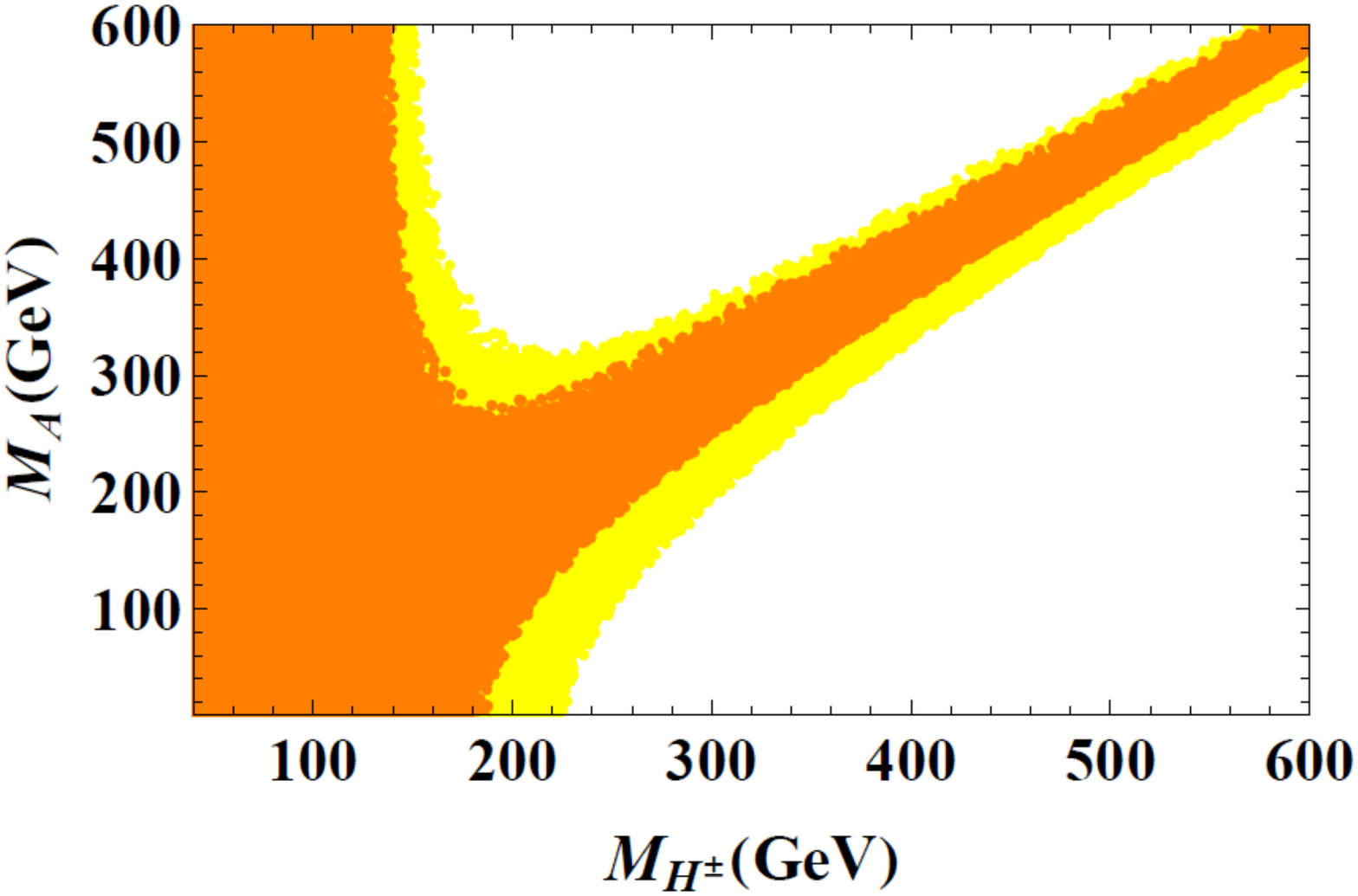}\includegraphics[scale=0.26]{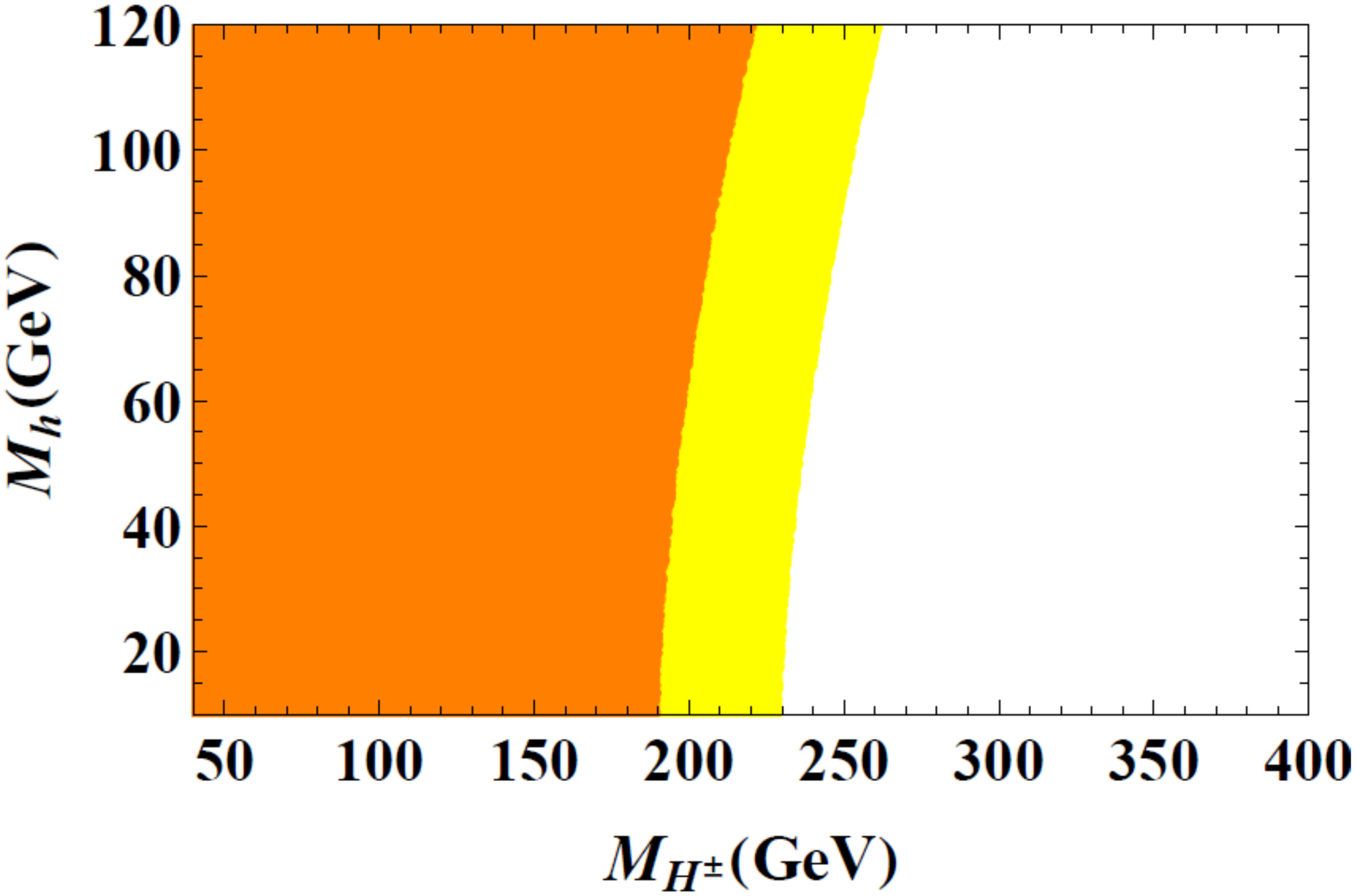}
\caption{One and two-sigma allowed region in the ($M_A,M_{H^\pm}$) plane for the case $M_H=125$ GeV (left) and in the ($M_h,M_{H^\pm}$) plane for $M_H=M_A=125$ GeV (right), from the oblique parameters S, T and U. The coupling $\cos\ta\in [0.8,1]$ and $M_h\in[10,120]$ GeV in both cases.}
\label{STU}
\end{figure}

The recent LHC discovery of a boson with mass of 125 GeV is the first direct hint of the electroweak symmetry breaking mechanism. Even if the current data suggest that the newly discovered particle is compatible with the SM Higgs, more and refined data is needed. However, we can now start putting indirect constraints on extended scalar sectors using the value of the Higgs mass. A simple example are the oblique parameters \cite{Baak:2012kk,LEPEWWG} as in Fig.\ref{STU}. An interesting possibility is to consider that the boson with mass of 125 GeV is the heaviest scalar $H$. Important constraints are derived when $M_{H^\pm},  \!\ M_A \geq 250$ GeV with $M_h\in[10,120]$ GeV; the oblique parameters require that the masses of the two bosons have to be quasi-degenerate. Another highly interesting possibility is to consider $M_H=M_A=125$ GeV \cite{Paper:2013rcs}, again with $M_h\in[10,120]$ GeV. In this case the oblique parameters require the presence of a charged Higgs just around the corner, approximately below the electroweak breaking scale 246 GeV. 

Future improvements of the bounds on neutral and charged Higgs, or perhaps their direct discovery at the LHC in the next few years might shed some light on the problem. In this paper I have presented an update of the analysis presented in \cite{Paper:2013rcs} using the latest data from the LHC. Other interesting possibilities are presented therein.

\section*{Acknowledgements}
I would like to thank the organizers of the EPS HEP 2013 conference for giving me the opportunity of presenting this work. I would also like to thank A. Pich and A. Celis for their collaboration on this project. This work has been supported in part by the Spanish Government and ERDF funds from the EU Commission [Grants No. FPA2011-23778, No. CSD2007-00042 (Consolider Project CPAN)] and by Generalitat
Valenciana under Grant No. PROMETEOII/2013/007.

\end{document}